\documentclass[%
twocolumn,
superscriptaddress,
nofootinbib,
 amsmath,amssymb,
 aps,
]{revtex4-2}

\usepackage{microtype}
\usepackage{graphicx}
\usepackage{dcolumn}
\usepackage{bm}

\usepackage{amssymb}
\usepackage{amsmath}
\usepackage{amsthm}
\usepackage{mathtools}
\usepackage[usenames,dvipsnames]{xcolor}
\usepackage{epsfig}
\usepackage{dcolumn}
\usepackage{tikz}
\usepackage{tikz-cd}
\usetikzlibrary{calc}
\usepackage{braket}
\usepackage{adjustbox}
\usepackage{hyperref}
\usepackage{verbatim}
\hypersetup{
	colorlinks=true,
	urlcolor=Maroon,
	linkcolor=RoyalBlue,
	citecolor=Maroon,
	bookmarks=true,
	pdftitle={Feynman Polytopes and the Tropical Geometry of UV and IR Divergences},
	pdfauthor={Nima Arkani-Hamed, Aaron Hillman, Sebastian Mizera},
	pdfdisplaydoctitle=true,
	pdfstartview=FitH,
 breaklinks=true
}

\definecolor{color2}{rgb}{0.368417, 0.506779, 0.709798}
\definecolor{color3}{rgb}{0.880722, 0.611041, 0.142051}
\definecolor{color5}{rgb}{0.560181, 0.691569, 0.194885}
\definecolor{color1}{rgb}{0.922526, 0.385626, 0.209179}
\definecolor{color6}{rgb}{0.528488, 0.470624, 0.701351}
\definecolor{color4}{rgb}{0.772079, 0.431554, 0.102387}

\usetikzlibrary{decorations.pathmorphing}
\tikzset{massless/.style={solid, line width=1.1pt}}
\tikzset{massive/.style={solid, line width=1.25pt, double}}
\tikzset{massive2/.style={solid, line width=0.9pt, double}}

\def \leq {\leqslant}
\def \geq {\geqslant}
\def \Newt {\mathrm{Newt}}

\def \Trop {\mathrm{Trop}}

\begin{document}

\title{A Subtraction Scheme for Feynman Integrals}

\author{Aaron Hillman}
\affiliation{%
Department of Physics, Jadwin Hall, Princeton University, NJ 08540, USA
}%
\affiliation{Walter Burke Institute for Theoretical Physics, California Institute of Technology, Pasadena, CA 91125}
\begin{abstract}
We present a subtraction scheme for ultraviolet (UV) divergent, infrared (IR) safe scalar Feynman integrals in dimensional regularization with any number of scales. This is done by the introduction of $u$-variables, which are a suitable generalization of dihedral coordinates on the open string moduli space  to Feynman integrals. The subtraction scheme furnishes subtraction terms which are products of lower loop Feynman integrals deformed by order $\epsilon$ powers of $u$-variables and deformations of the degree of divergence. The result is a canonical and algorithmic prescription to express the Feynman integral as a sum of convergent integrals dressed with inverse powers of $\epsilon$.
\end{abstract}

\maketitle
\section{Introduction}

With the challenges multi-loop and multi-scale Feynman integrals present, it is valuable to have a variety of strategies for evaluation, in particular numerical evaluation. In order to find a convergent presentation of a Feynman integral in dimensional regularization without analytic continuation, a subtraction scheme is required. For a generic multi-scale Feynman integral in dimensional regularization, there is no canonical subtraction scheme, in particular not one leveraging the factorization properties associated with ultraviolet (UV) divergences. A more general method applicable for a wide class of integrals can be employed \cite{nilsson2010mellin}, but proliferates in complexity and obfuscates the factorization structure associated with the resummation of divergences. The need for a subtraction scheme is not only a practical hurdle to calculation but also an element of the hurdle in understanding the resummation of divergences directly from an on-shell point of view. We therefore believe this is a problem deserving of a conceptually clear and practically useful resolution.\\ We present a subtraction scheme for calculating the Laurent expansion of UV divergent, infrared (IR) safe Feynman integrals in dimensional regularization. The scheme is analogous to that furnishing the $\alpha'$-expansion of open string amplitudes \cite{Brown:2019wna} and requires the introduction of $u$-variables, which are a suitable generalization of dihedral coordinates \cite{Brown:2009qja} (i.e. cross-ratios) on the open-string moduli space to Feynman integrals\footnote{See \cite{Brown:2015fyf}, \S 5.4 for a related construction.}. These variables furnish a binary realization of a triangulation of the integrand's associated tropical fan, which we call the BPHZ triangulation. The $u$-variables naturally motivate a class of generalized Feynman integrals with integrands supplemented by factors of $u$'s raised to powers which are positive integers times $\epsilon$. The subtraction terms are manifestly products of lower-loop Feynman integrals in  this class, whose Laurent expansions are themselves computed at lower loop order in the same way.\\
In this work we articulate the essential non-trivial facts undergirding the subtraction scheme and illustrate in a few examples. For those uninterested in the non-trivial properties of $u$-variables more broadly, the scheme can be read as a list of finite integrals dressed with inverse powers of $\epsilon$ which can be numerically integrated to compute the Laurent expansion of a Feynman integral.
\section{Feynman Polytopes}
The integrals whose Laurent expansion we are interested in calculating are 
\begin{equation}
    \label{eq:feynint}
    I_G = \int \omega_G \frac{\prod\limits\alpha_e}{\mathcal{U}^{D/2}}\left(\frac{\mathcal{U}}{\mathcal{F}} \right)^{d_G}
\end{equation}
where $d_G = E-LD/2$ and the contribution to the amplitude comes with a factor $\Gamma(d_G)$ in front. We have the usual Symanzik polynomials defined in terms of the graph $G$'s spanning one-trees and two-trees $T^1$ and $T^2$ respectively
\begin{align}
    \mathcal{U} &= \sum\limits_{T^1} \prod\limits_{e\notin T^1}\alpha_e\\
    \mathcal{F} &= \left(\sum\limits_e m_e^2\alpha_e\right)\mathcal{U}+\sum\limits_{T^2} (p_{T^2})^2\prod\limits_{e\notin T^2}\alpha_e
\end{align}
and momentum $p_{T^2}$ flowing across the two-tree. We also have the standard projective meausure
\begin{equation}
    \label{eq:projmeasure}
    \omega_G  =\sum\limits_e (-1)^e d\log(\alpha_1)\dots \widehat{\log( \alpha_e)}\dots d\log(\alpha_E)
\end{equation}
where the hat denotes omission. We have written the above in Euclidean signature for simplicity, but all results hold in Lorentzian signature in the absence of IR singularities, with contour deformations also needing to be performed for kinematics above threshold. 
\paragraph*{\bf Feynman polytopes} The natural way to understand the asymptotic behavior of these integrals is in the language of tropical fans and dual Newton polytopes. For a subset of Lorentzian kinematics, the facet inequalities which define 
\begin{equation}
    \mathbf{S}_G := \Newt[\mathcal{U}]\oplus \Newt[\mathcal{F}]
\end{equation}
were described in \cite{Arkani-Hamed:2022cqe}. For the case in which the Feynman polytopes are generalized permutahedra, they were already understood in \cite{Schultka:2018nrs}. The inequalities and compatibility condition are described in terms of a single function 
\begin{equation}
\label{eq:zdef}
    z_\gamma := \begin{cases}
      \mathcal{F}_{G/\gamma} \neq 0  & 2L_\gamma\\
        \mathcal{F}_{G/\gamma} = 0 &  2L_\gamma+1
    \end{cases}
\end{equation}
In which case we can describe in $E$ homogenous coordinates the 
\begin{equation}
\label{eq:SGineqs}
   \mathbf{S}_G : \hspace{4mm }a_G = \{2L_G+1\} \text{   and   } \{ a_\gamma \geq z_\gamma \} 
\end{equation}
where 
\begin{align}
    a_\gamma &= \sum\limits_{e \in \gamma} a_e\\
    \alpha_\gamma &= \sum\limits_{e \in \gamma} \alpha_e
\end{align}
The $a_e$ variables are dual to the logarithms of the $\alpha_e$, in particular the facet normals of the polytope are scaling directions of the logarithms of the $\alpha_e$. These inequalities are tied to the factorization properties of the Symanzik polynomials associated with taking the leading behavior as the edges in a subgraph are rescaled by a common parameter. Ultraviolet divergences are associated with the first inequality of (\ref{eq:zdef}) and we will utilize the associated factorization property
\begin{align}
    \mathcal{U}_G &= \mathcal{U}_{\gamma}\times \mathcal{U}_{G/\gamma}+\mathcal{O}(\alpha_\gamma^{L_\gamma+1})\\
    \mathcal{F}_G &= \mathcal{U}_{\gamma}\times \mathcal{F}_{G/\gamma}+\mathcal{O}(\alpha_\gamma^{L_\gamma+1})
\end{align}
Finally the compatibility condition for two facets is 
\begin{equation}
    z_{\gamma_1} +z_{\gamma_2} \geq z_{\gamma_1\cup \gamma_2}+z_{\gamma_1 \cap\gamma_2}  
\end{equation}
this is consistent with but in the opposite direction of the supermodularity condition for  inequalities of the form in (\ref{eq:SGineqs}) to cut out a generalized permutahedron.  Compatibility is therefore saturation of the above inequality in the generalized permutahedron case.\\
The $u$-variables and associated binary realization we will describe later on will only apply with kinematics for which $\mathbf{S}_G$ is a generalized permutahedron, in particular all subgraphs obey
\begin{equation}
    z_{\gamma_1} +z_{\gamma_2} \leq z_{\gamma_1\cup \gamma_2}+z_{\gamma_1 \cap\gamma_2}  
\end{equation}
We emphasize though, that despite this requirement for the $u$-variables to describe the full geometry, the application of the subtraction scheme is more general. This is because we will only ever need subtractions associated to divergent subgraphs. Therefore, the condition for applicability is that the divergent part of the fan is identical to that of a graph which is a generalized permutahedron. Therefore, the subtraction scheme will apply for any infrared safe graph (and even some infrared divergent graphs).
\section{\texorpdfstring{$u$}{u}-Variables}
In this section we describe the variables which will facilitate the subtraction scheme. These are the analogue of the conformal-cross ratios on the open string world-sheet \cite{Koba:1969kh}, which were generalized to finite-type cluster algebras \cite{Arkani-Hamed:2019mrd, Arkani-Hamed:2019plo} and other binary geometries \cite{He:2020onr}, but for our problem. This generalization is natural in the language of Newton polytopes and tropical fans employed previously in \cite{Arkani-Hamed:2022cqe}.
\paragraph*{\bf Tropical construction } The facet rays are associated with one-vertex irreducible (1VI) subgraphs $\gamma$ and to each facet we associate a $u_\gamma$. The most efficient way to construct $u_\gamma$ is via the application of the tropical condition
\begin{equation}
\label{eq:tropu}
    \Trop[u_\gamma](r_{\gamma'}) = \delta_{\gamma \gamma'}
\end{equation}
to the Ansatz 
\begin{equation}
    u_\gamma \big |_G = \prod\limits_{\Gamma \supseteq \gamma} \left(\frac{\alpha_\Gamma}{\alpha_G} \right)^{p_\Gamma}
\end{equation}
where the product is over 1VI $\Gamma$'s which contain $\gamma$ in addition to $\gamma$ itself. 
 The evaluation at $G$ denotes defining $u_\gamma$ with respect to total graph $G$ and we will drop this notation when the context is clear. We can straightforwardly derive the recursive solution to (\ref{eq:tropu}) by observing that 
\begin{equation}
    \Trop[\alpha_\Gamma](r_{\gamma'}) = \begin{cases}
        1 & \Gamma \subseteq \gamma'\\
        0 & \text{ else}
    \end{cases}
\end{equation}
This can be easily seen noting the rays have entries 1 or 0 and \text{Trop} is the \text{Min} over the components. Note that $\Trop[\alpha_G] = 0$ as there is always a zero entry for any ray. When applied to the tropicalization of the Ansatz, we find
\begin{equation}
    \Trop[u_\gamma] = \sum\limits_{\Gamma \supseteq \gamma} p_\Gamma \Trop[\alpha_\Gamma]
\end{equation}
yields $p_\gamma = 1$ and
\begin{equation}
    \label{eq:precursion}
    p_\Gamma = -\sum\limits_{\Gamma \supset \Gamma' \supseteq \gamma}p_{\Gamma'}
\end{equation}
The power of $p_\Gamma$ is fixed by the powers of all the graphs it contains. This is naturally graded by loop number, and so one may proceed determining the powers $p_\Gamma$ by increasing loop number $L_\Gamma \geq L_\gamma$ starting with $p_\gamma = 1$ which is always the case.
\paragraph*{\bf Properties of the $u$'s} The first non-trivial property of the $u$'s which we will highlight is that they furnish a binary realization not, strictly speaking, of the Feynman polytope but of what we may call a triangulation of its tropical fan. We can only have $u$-variables associated to a simplicial fan and therefore a simple polytope, and Feynman polytopes are not simple in general. But a triangulation of the fan is simple, and corresponds in the dual Newton polytope to shifting constants in the facet inequalities so that the rays are identical, but now only $d$ facets ever meet at a vertex in $d$ dimensions. Equivalently, this introduces additional constraints for compatibility. The particular triangulation which this construction lands on is in fact the BPHZ triangulation, generalized to all subgraphs. That is, two subgraphs are only compatible if they are nested or disjoint. In this sense, the BPHZ forest formula is not merely a particular solution to Bogolyubov's recursion, but completely canonical.\\
We can state the binary property of the $u_\gamma$'s by stating their limiting behavior upon shrinking some other facet subgraph $\gamma'$. At the level of the $\alpha_e$ variables, this means only keeping in each factor $\alpha_\Gamma$ the leading $\alpha_e$ i.e. those $e$ which are not in $\gamma'$. The binary property is then stated as
 \begin{equation}
 \label{eq:binaryconditions}
     u_{\gamma_j} \to \begin{cases}
         u_{\gamma_j} \big |_{\gamma_i} & \gamma_j \subset \gamma_i \\
         u_{\gamma_j } \big |_{G/\gamma_i} & \gamma_i \subset \gamma_j \text{ or } \gamma_j, \gamma_i \text{ disjoint } \\
         1 & \text{ otherwise } \\
     \end{cases}
 \end{equation}
 where incompatible means not satisfying either of the first two criteria i.e. not being nested or disjoint. This is called a binary realization because all facets are probed by a $u_\gamma$ going to 0 and its incompatible facets going to 1. Moreover, any set of $E-1$ compatible $u_\gamma$'s constitutes a good set of integration variables which vary over the hypercube $[0, 1]^{E-1}$. These properties of the $u_\gamma$'s will be crucial for the simplicity of our renormalization scheme.\\
 The other non-trivial property the $u_\gamma$'s posses is the inversion property 
 \begin{equation}
      \label{eq:alphainvert}
     \frac{\alpha_e}{\alpha_G}  = \prod\limits_\gamma u_\gamma^{\gamma | e}
 \end{equation}
  where $\gamma | e$ simple denotes whether $e$ is an edge in $\gamma$ or not i.e.
 \begin{equation}
     \gamma | e = \begin{cases}
         1 & e \in \gamma \\
         0 & e \notin \gamma
     \end{cases}
 \end{equation}
 The final property we could hope to have are $u$-equations. It appears that the $u_\gamma$'s do not obey perfect $u$-equations save for exceptional cases where the geometry is associated to a finite-type cluster algebra, such as in the case of the sunrise graph which corresponds to the two-dimensional cyclohedron $B_2$.\\
 Note that by the projective invariance of the integrand we can express the integrand purely in terms of $u_\gamma$'s using the inversion formula. We will denote
 \begin{equation}
     U_e = \prod\limits_\gamma u_\gamma^{\gamma | e}
 \end{equation}
It will be useful later to introduce the following $\epsilon$-deformed generalization of the Feynman integral
\begin{equation}
\label{eq:genint}
     I_G(\{ p_\gamma \}, \delta d_G) = \int \omega_G \left(\prod\limits_\gamma u_\gamma^{ p_\gamma}\right) \frac{\prod\limits U_e}{\mathcal{U}(U)^{D/2}}\left(\frac{\mathcal{U}(U)}{\mathcal{F}(U)} \right)^{d_G+\delta d_G}
 \end{equation}
 where this equals the Feynman integral we are interested in for $ p_\gamma = 0$ and $\delta d_G= 0$ and the argument $U$ means replacing $\alpha_e$ with $U_e$. Projective invariance ensures that this is a rewriting of the integrand. Our subtraction terms will always be equal to integrals in this class for $p_\gamma$ which are positive integers times $\epsilon$. It is important to emphasize that when directly working in the underlying $\alpha_e$ variables and accounting for a non-trivial $\delta d_G$, the function which is raised to the shift $\delta d_G$ is the projectively invariant ratio
 \begin{equation}
     \left(\frac{\alpha_G \mathcal{U}(\alpha_e)}{\mathcal{F}(\alpha_e)}\right)^{\delta d_G}
 \end{equation}
\paragraph*{\bf Parachute example}
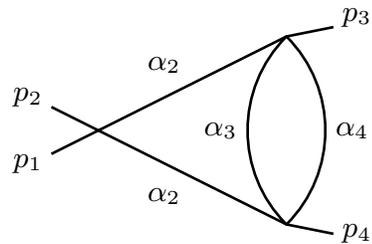
\begin{figure}
     \centering
     \scalebox{1.25}{\begin{tikzpicture}
     \draw[thick] (0, 0) -- (2, 1) to [ out = 315, in = 45] (2, -1) -- (0, 0);
     \draw[thick] (2, -1) to [out = 135, in = 225] (2, 1);
     \draw[thick] (-.5, .25) -- (0, 0) -- (-.5, -.25);
     \draw[thick] (2, 1.0) to (2.5, 1.1);
     \draw[thick] (2, -1.0) to (2.5, -1.1);
     \node at (.7, -.7) {$\alpha_2$};
     \node at (.7, .7) {$\alpha_2$};
     \node at (1.3, 0) {$\alpha_3$};
     \node at (2.7, 0) {$\alpha_4$};
     \node at (2.75, 1.2) {$p_3$};
     \node at (2.75, -1.1) {$p_4$};
     \node at (-.75, -.35) {$p_1$};
     \node at (-.75, .35) {$p_2$};
     \end{tikzpicture}}
     \caption{Two-loop $\lambda\phi^4$ graph with generic kinematics. We refer to the graph with zero external kinematics and $m_e^2 = 1$ as a ``UV" graph, as these are the effective kinematics appearing associated with a shrinking subgraph. }
     \label{fig:parachute}
 \end{figure}
 We can illustrate these properties in a simple example, the parachute graph depicted in figure \ref{fig:parachute}. In this case, we find the variables
 \begin{align*}
\hspace{10mm}u_1 = \frac{\alpha_{1}\alpha_G}{\alpha_{123}\alpha_{124}} \hspace{10mm}  u_2 = \frac{\alpha_{2}\alpha_G}{\alpha_{123}\alpha_{124}} \hspace{10mm}\\[3mm]
    \hspace{7mm}u_3 = \frac{\alpha_{3}\alpha_G}{\alpha_{34}\alpha_{123}} \hspace{10mm}  u_4 = \frac{\alpha_{4}\alpha_G}{\alpha_{34}\alpha_{124}} \hspace{13mm}\\[3mm]
    \hspace{5mm}  u_{123} =  \frac{\alpha_{123}}{\alpha_G}  \hspace{10mm} u_{124} =   \frac{\alpha_{124}}{\alpha_G}  \hspace{15mm} \\[3mm]
     u_{34} = \frac{\alpha_{34}}{\alpha_G}\hspace{28mm}
 \end{align*}
Using these variables equation (\ref{eq:alphainvert}) is readily verified.
\begin{table}
    \centering
    \begin{tabular}{|c|c|c|c|c|c|c|c|}
       \hline \rule{0pt}{3ex}  & $u_1$ & $u_2$ & $u_3$ & $u_4$ & $u_{34}$  & $u_{123}$ &  $u_{124}$\\[2ex]
         \hline \rule{0pt}{3ex}  $u_1$ & $\frac{\alpha_{1}\alpha_G}{\alpha_{123}\alpha_{124}}$  & $\frac{\alpha_2\alpha_{234}}{\alpha_{23}\alpha_{24}} $ & $\frac{\alpha_3\alpha_{234}}{\alpha_{23}\alpha_{34}} $ & $\frac{\alpha_4\alpha_{234}}{\alpha_{24}\alpha_{34}} $ & $\frac{\alpha_{34}}{\alpha_{234}} $  & $\frac{\alpha_{23}}{\alpha_{234}} $ &$\frac{\alpha_{24}}{\alpha_{234}} $ \\[2ex]
        \hline  \rule{0pt}{3ex} $u_2$ &  $\frac{\alpha_1\alpha_{134}}{\alpha_{13}\alpha_{14}}$ & $\frac{\alpha_{2}\alpha_G}{\alpha_{123}\alpha_{124}}$  & $\frac{\alpha_3\alpha_{134}}{\alpha_{13}\alpha_{34}} $ & $\frac{\alpha_4\alpha_{134}}{\alpha_{14}\alpha_{34}} $ & $\frac{\alpha_{34}}{\alpha_{134}} $  & $\frac{\alpha_{13}}{\alpha_{134}} $ & $\frac{\alpha_{14}}{\alpha_{134}} $ \\[2ex]
        \hline \rule{0pt}{3ex} $u_3$ & $\frac{\alpha_1}{\alpha_{12}} $ & $\frac{\alpha_2}{\alpha_{12}} $ & $\frac{\alpha_{3}\alpha_G}{\alpha_{34}\alpha_{123}}$ & 1 & $\frac{\alpha_4}{\alpha_{124}} $  & $\frac{\alpha_{12}}{\alpha_{124}} $ & 1 \\[2ex]
         \hline \rule{0pt}{3ex} $u_4$ &$\frac{\alpha_1}{\alpha_{12}} $ & $\frac{\alpha_2}{\alpha_{12}} $ & 1 & $\frac{\alpha_{4}\alpha_G}{\alpha_{34}\alpha_{124}}$ & $\frac{\alpha_3}{\alpha_{123}} $ & 1 & $\frac{\alpha_{12}}{\alpha_{123}} $ \\[2ex]
        \hline \rule{0pt}{3ex} $u_{34}$ & $\frac{\alpha_1}{\alpha_{12}} $ & $\frac{\alpha_2}{\alpha_{12}} $ & $\frac{\alpha_3}{\alpha_{34}}$ & $\frac{\alpha_4}{\alpha_{34}}$ & $\frac{\alpha_{34}}{\alpha_G}$ & 1 & 1 \\[2ex]
        \hline \rule{0pt}{3ex} $u_{123}$ & $\frac{\alpha_1}{\alpha_{123}} $ & $\frac{\alpha_2}{\alpha_{123}} $ & $\frac{\alpha_3}{\alpha_{123}} $ & 1 & 1 & $\frac{\alpha_{123}}{\alpha_G}$  & 1 \\[2ex]
        \hline \rule{0pt}{3ex} $u_{124}$ & $\frac{\alpha_1}{\alpha_{124}} $ & $\frac{\alpha_2}{\alpha_{124}} $ & 1 & $\frac{\alpha_4}{\alpha_{124}} $ & 1 & 1  & $\frac{\alpha_{124}}{\alpha_G}$ \\[2ex]
        \hline
    \end{tabular}
    \caption{The limit of the $u_{\gamma}$ variable in a given column as the $u_{\gamma'}$ in row goes to zero. Notice that the simplifications reflect the compatibility described in (\ref{eq:binaryconditions}). These $u_\gamma$ therefore furnish a factorizing and binary realization of the geometry.}
    \label{tab:parachutecompat}
\end{table}
The table of co-dimension one limits is present in table \ref{tab:parachutecompat}.
\section{Subtraction Scheme}
Subtraction terms are associated with simplicial cones in our triangulated fan, or dually, faces. As stated above, these faces are equivalent to BPHZ forests, i.e. sets of subgraphs which are all pairwise nested or disjoint. We denote these by $F$. We will utilize the shorthand $\cdot/F$ where whatever is being modded by $F$ is understood as simplified on the support of all subgraphs in $F$ shrinking. In particular, when constructing a subtraction term, any member $\Gamma$ of $F$ and its corresponding $u_\Gamma$ are simplified on the support of all other $u_{\Gamma'} \to 0$ and we denote this as $u_{\Gamma/F}$ i.e. keeping only terms leading as the other members of $F$ go to zero. Precisely this means dropping the subleading $\alpha_e$ in the $\alpha_\Gamma$. This is only an unambiguous operation for compatible facets. We are now equipped to state the subtraction terms.
\paragraph*{\bf Subtraction term} The substraction term associated to a single $F = \{ \Gamma_i\}$ for the integral (\ref{eq:genint}) is
\begin{equation}
\label{eq:sf}
    S_F = \int \Omega_F
\end{equation}
where
\begin{multline}
\label{eq:of}
    \Omega_F = \bigwedge\limits_i(u_{\Gamma_i/F})^{d_{\Gamma_i}}d\log(u_{\Gamma_i/F})\\ \wedge\left[\omega_{G/F}\frac{\prod\limits_e^{G/F} U_{e}}{\mathcal{U}_{G/F}^{D/2}}\left(\frac{\mathcal{U}_{G/F}}{\mathcal{F}_{G/F}} \right)^{d_{G/F}+\delta d_{G/F}}  u_{\gamma_j/F}^{p_{\gamma_j}+\delta p_{\gamma_j/F}}\right]\\\bigwedge\limits_i\left[\omega_{\Gamma_i/F}\frac{\prod\limits_e^{\Gamma_i/F} U_{e}}{\mathcal{U}_{\Gamma_i/F}^{D/2}} u_{\gamma_k/F}^{p_{\gamma_k}+ \delta p_{\gamma_k/F}} \right]
\end{multline}
where a product over the $j$'s, and $k$'s inside of each factorized subgraph expression is implicit and the $\delta p_{y_j/F}$ are integer shifts of the $u_{\gamma_j/F}$ exponents which depend on $F$
\begin{equation}
    \label{eq:deltap}
     \delta p_{\gamma_j/F} = (L_{\gamma_j}-L_{\gamma_j/F})\epsilon
\end{equation}
Where $\gamma_j$ denotes the graph in $G$ which reduced to $\gamma_j/F$, a subgraph either of $G/F$ or some $\Gamma_i/F$. We have simply stated this subtraction term, but it can be calculated via a modified residue prescription as in \cite{Brown:2019wna}.\\
The integration in the $u_{\Gamma_i/F}$ is trivial and simply produces inverse powers of $d_{\Gamma_i}$ and the remaining integrals have all factorized and are merely lower-loop integrals of the type in (\ref{eq:genint}). In particular
\begin{equation}
    \label{eq:sfintegrated}
    S_F = \frac{1}{\prod\limits_i d_{\Gamma_i}}I_{G/F}(\{p_{\gamma}'\}_{G/F}, \delta d_{G/F})\prod\limits_i I_{\Gamma_i/F}(\{p_{\gamma}' \}_{\gamma_i/F})
\end{equation}
where the $p_\gamma'$ are the shifted exponents in (\ref{eq:of}) and
\begin{equation}
\label{eq:dodshift}
    \delta d_{G/F} = d_G-d_{G/F}
\end{equation}
The $I_{\Gamma_i/F}$ have specific kinematics: zero external kinematics and $m_e^2 = 1$. In this sense, the appropriate subtraction terms are built out of products of integrals dominated by the ultraviolet for the internal legs, where the external scales are effectively shut off.
\paragraph*{\bf Subtraction scheme}
It remains to specify precisely how we subtract to get a convergent integral. The correct procedure is to subtract all $F$ terms with alternating sign $(-1)^{|F|}$, with $|F|$ the number of subgraphs in $F$. This is simply inclusion exclusion: the subtraction associated with a given $F$ contains subdivergences associated with all $F'$'s of which $F$ is a subset. In view of this, the alternating sign ensures no double counting. In particular we have the renormalized integrand
\begin{equation}
\label{eq:renform}
    \tilde \Omega_G = \Omega_G+\sum\limits_{F} (-1)^{|F|}\Omega_F
\end{equation}
Which is finite as $\epsilon \to 0$ and can be expanded order-by-order in $\epsilon$ at integrand level. It is clear combining (\ref{eq:sfintegrated}) and (\ref{eq:renform}) that
\begin{equation}
\label{eq:IGexpress}
    I_G =  -\sum\limits_F (-1)^{|F|} S_F+\int \tilde \Omega_G
\end{equation}
 Moreover, everything in $S_F$ is computed at lower loop order. We emphasize that (\ref{eq:IGexpress}) combined with (\ref{eq:sfintegrated}) furnishes a manifestly convergent representation of the Feynman integral.
\section{Examples}
\paragraph*{\bf Parachute} First we consider the parachute. We will need the parachute for zero external kinematics and $m_e^2 = 1$ when computing higher loop graphs later on. First we will consider the subtraction for totally generic kinematics.
\begin{multline}
\label{eq:parachutesubgen}
    S_{34} = \int\limits u_{34}^{\epsilon}d\log(u_{34}) d\log(\alpha_1/\alpha_2)d\log(\alpha_3/\alpha_4)\\
    \times \frac{\alpha_1\alpha_2\alpha_3\alpha_4}{(\alpha_1+\alpha_2)^2(\alpha_3+\alpha_4)^2}
   \\ \times \left(\frac{(\alpha_1+\alpha_2)^2}{s \alpha_1\alpha_2+(m_1^2\alpha_1+m_2^2\alpha_2)(\alpha_1+\alpha_2)}  \right)^{2\epsilon}
\end{multline}
with $s = (p_1+p_2)^2$ and a choice of chart with e.g. $\alpha_1 \to 1$ and the remaining variables integrated from zero to infinity is implicit (and is implicit in the remaining integrals). This is equal to
\begin{multline}
    S_{34} = \frac{1}{\epsilon}\int d\log(\alpha_1/\alpha_2) \frac{\alpha_1\alpha_2}{(\alpha_1+\alpha_2)^2}\\
    \times \left(\frac{(\alpha_1+\alpha_2)^2}{s \alpha_1\alpha_2+(m_1^2\alpha_1+m_2^2\alpha_2)(\alpha_1+\alpha_2)}  \right)^{2\epsilon}  
\end{multline}
where two integrations have been performed, leaving a one-dimensional integral. This is simply a bubble with the appropriate resulting kinematics from shrinking $\gamma_{34}$ and with $\delta d_{G/F} = \epsilon$ in accordance with  (\ref{eq:dodshift}). This integral is convergent and expandable order by order in $\epsilon$ at integrand level. In the limit of zero external kinematics and $m_e^2 = 1$, we in fact have that $S_{34} = \frac{1}{\epsilon}$. This yields the expandable expression for the Feynman integral with generic kinematics
\begin{equation}
    \label{eq:parachutegen}
    I_{\text{par}} = S_{34}+\int(\Omega_G-\Omega_{34})
\end{equation}
where $\Omega_{34}$ is simply the integrand in (\ref{eq:parachutesubgen}). We state this explicitly in $\alpha_e$ parameters for the case of zero external kinematics and $m_e^2 = 1$
\begin{multline}
    \label{eq:parachuteUVkin}
    I_{\text{par}}^{\text{UV}} = \frac{1}{\epsilon} +\int d\alpha_2 d\alpha_3 d\alpha_4 \bigg[ \frac{(1+\alpha_2+\alpha_3+\alpha_4)^{-2\epsilon}}{[(\alpha_1+\alpha_2)(\alpha_3+\alpha_4)+\alpha_3\alpha_4 ]^{2-\epsilon}}\\
  - \left(\frac{\alpha_3+\alpha_4}{1+\alpha_2+\alpha_3+\alpha_4} \right)^{\epsilon}\times\\ \frac{1}{(1+\alpha_2+\alpha_3+\alpha_4)(\alpha_1+\alpha_2)(\alpha_3+\alpha_4)^2}  \bigg] 
\end{multline}
\begin{figure}
    \centering
    \scalebox{.9}{\begin{tikzpicture}
        \coordinate (v1) at (-3, 0);
        \coordinate (v2) at (3, 0);
        \coordinate (v3) at (0, 0);
        \coordinate (eL1) at (-3.6, .4);
        \coordinate (eL2) at (-3.5, -.4);
        \coordinate (eR1) at (3.5, .4);
        \coordinate (eR2) at (3.5, -.4);
        \draw[very thick] (v1) to[ out=70, in=110] (v3) to[very thick, out=250, in=-70] (v1);
        \draw[very thick] (v3) to[ out=70, in=110] (v2) to[very thick, out=250, in=-70] (v3);
        \draw[very thick] (eL1) to (v1) to[very thick] (eL2);
        \draw[very thick] (eR1) to  (v2) to[very thick] (eR2);
        \node at ($(v1)+(1.5, 1.1)$) { \large $\alpha_2$};
        \node at ($(v1)+(1.5, -1.1)$) {\large $\alpha_1$};
        \node at ($(v2)+(-1.5, 1.1)$) { \large $\alpha_3$};
        \node at ($(v2)+(-1.5, -1.1)$) {\large $\alpha_4$};
    \end{tikzpicture}}
    \caption{Product of bubbles in $\phi^4$ with $m_e^2 = 0$}
    \label{fig:bubbleprod}
\end{figure}
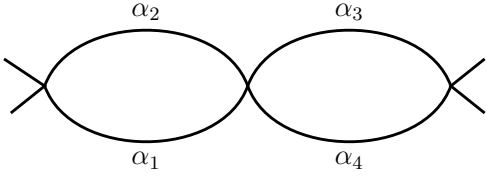
\begin{figure}
    \centering
    \scalebox{.75}{\begin{tikzpicture}
        \coordinate (v1) at (-4, 0);
        \coordinate (v2) at (4, 0);
        \coordinate (v3) at (0, 2);
        \coordinate (v4) at (0, -2);
        \coordinate (eL1) at (-5, .5);
        \coordinate (eL2) at (-5, -.5);
        \coordinate (eR1) at (5, .5);
        \coordinate (eR2) at (5, -.5);
        \draw[very thick] (v1) to[very thick] (v3) to[very thick] (v2) to[very thick] (v4) to[very thick] (v1);
        \draw[very thick] (v3) to [very thick, out = -30, in = 30] (v4) to [very thick, out = 150, in = 210] (v3);
        \draw[very thick] (eL1) to[very thick] (v1) to[very thick] (eL2);
        \draw[very thick] (eR1) to[very thick] (v2) to[very thick] (eR2);
        \node at ($(v1)+(1.3, 1.1)$) { \large $\alpha_2$};
        \node at ($(v1)+(1.3, -1.1)$) {\large $\alpha_1$};
        \node at ($(v2)+(-1.3, 1.1)$) { \large $\alpha_3$};
        \node at ($(v2)+(-1.3, -1.1)$) {\large $\alpha_4$};
        \node at ($(v1)+(2.6, 0)$) { \large$\alpha_5$};
        \node at ($(v2)+(-2.6, 0)$) {\large $\alpha_6$};
    \end{tikzpicture}}
    \caption{Three loop single scale graph in $\phi^4$ with $m_e^2 = 0$}
    \label{fig:eyegraph}
\end{figure}
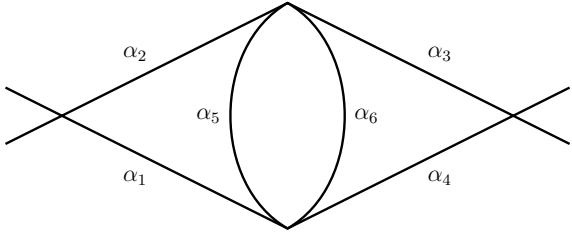
\paragraph*{\bf Bubble product} The product of bubbles is of course obtainable from the bubble alone, but not necessarily for integrals in the class (\ref{eq:genint}) where the presence of the $u_\gamma$ can entangle the variables in each bubble. In this case, in order to compute the Laurent expansion there is simply a subtraction term for each bubble in figure \ref{fig:bubbleprod}, $\gamma_{12}$ and $\gamma_{34}$. The relevant $u$-variables are
\begin{equation}
    \label{eq:bubbleprodus}
    u_{12} = \frac{\alpha_{12}}{\alpha_G}\hspace{10mm}u_{34} = \frac{\alpha_{34}}{\alpha_G}
\end{equation}
It is worth noting that they are obviously incompatible degenerations by the binary property.
We have the subtraction term
\begin{multline}
    S_{12} = \int\omega_{12} u_{12}^{\epsilon+p_{\gamma_{12}}}d\log(u_{12})\frac{U_1 U_2}{\mathcal{U}_{12}(U)^{2-\epsilon}}\\ 
    \times\omega_{34}\frac{U_3 U_4}{\mathcal{U}_{34}(U)^{2-\epsilon}}\left(\frac{\mathcal{U}_{34}(U)}{\mathcal{F}_{34}(U)} \right)^{2\epsilon+\delta d_G}
\end{multline}
and the equivalent for $\gamma_{34}$. This subtraction term is 
\begin{equation}
    S_{12} = \frac{1}{\epsilon+p_{\gamma_{12}}}\frac{\Gamma(1-\delta d_G-2\epsilon)^2}{\Gamma(1-2\delta d_G-4\epsilon)}
\end{equation}
Therefore we have 
\begin{multline}
    \label{eq:bubbxbubb}
    I_{\text{bubb}\times \text{bubb}}(\{p_{12} , p_{34}  \}, \delta d_G ) =\int (\Omega_G-\Omega_{12}-\Omega_{34})\\+ \left(\frac{1}{\epsilon+p_{\gamma_{12}}}+\frac{1}{\epsilon+p_{\gamma_{34}}}\right)\frac{\Gamma(1-\delta d_G-2\epsilon)^2}{\Gamma(1-2\delta d_G-4\epsilon)}
\end{multline}
\paragraph*{\bf Eye graph} Finally we consider the eye graph depicted in figure (\ref{fig:eyegraph}). This will require five subtraction terms yielding the subtracted integrand
\begin{equation}
    \tilde \Omega_G = \Omega_G-\Omega_{1256}-\Omega_{3456}-\Omega_{56}+\Omega_{56, 1256}+\Omega_{56, 3456}
\end{equation}
with the mutually degenarating subgraphs in each $F$ set off by commas. The relevant integrated values are
\begin{align}
    &\int \Omega_{1256} = \frac{1}{2\epsilon}I_{\text{bubb}}(\delta d_G = 2\epsilon)I_{\text{par}}^{\text{UV}}\\
    & \int \Omega_{56} = \frac{1}{\epsilon}I_{\text{bubb}\times \text{bubb}}(\{p_{12} = \epsilon, p_{34} = \epsilon \}, \delta d_G = \epsilon)\\
    &\int \Omega_{56, 1256} =  \frac{1}{2\epsilon}\frac{1}{\epsilon}I_{\text{bubb}}(\delta d_G = 2\epsilon)
\end{align}
with the rest of the terms attained by symmetry. We have the massless bubble with $s = 1$ and $\delta d_G = 2\epsilon$ in accordance with our prescription in (\ref{eq:of})
\begin{equation}
    \label{eq:shiftbubb}
    I_{\text{bubb}}(\delta d_G = 2\epsilon) = \frac{\Gamma(1-3\epsilon)^2}{\Gamma(2-6\epsilon)}
\end{equation}
The integral in the second line is merely that of a product of two bubbles $\gamma_{12}$ and $\gamma_{34}$ but with powers $p_{124} = p_{34} = \epsilon$ and $\delta d_G = \epsilon$. In particular, we computed its Laurent expansion above.
\begin{multline}
    \label{eq:eyefinal}
    I_G =\frac{\Gamma(1-\epsilon)^4}{\Gamma(1-4\epsilon)\epsilon^2}\left(1+7\epsilon+31\epsilon^2+(103+36\zeta(3))\epsilon^3 +\dots\right)
\end{multline}
In the single-scale case we verify that numerical integration is consistent with the result \cite{Chetyrkin:1980pr}.\\
None of the analysis above was sensitive to the kinematics of the three-loop graph other than the absence of infrared divergences and can be repeated for arbitrary mass scales as in the parachute case above.
\section{Outlook}
The above scheme is a canonical solution to the problem of computing the $\epsilon$-expansion of multi-scale Feynman integral with UV subdivergences, which we plan to implement as part of a package in upcoming work \cite{AHHM}. In addition to practical applicability, the scheme invites avenues for more formal exploration. We have not, for instance, connected this scheme to a renormalization scheme in the conventional sense, associating specific integrals with contributions to counter terms in renormalized perturbation theory. This may be interesting to explore. More broadly, we hope that this scheme may open a door to a clear understanding of renormalization directly in the space in which we calculate observables.\\
The $u$-variables introduced in order to furnish the subtraction scheme are interesting in their own right. It would be interesting to explore the existence of $u$-equations for these variables along the lines of \cite{He:2020onr}. With such $u$-equations and their solutions, one could define Feynman integrals intrinsically in terms of $u$-variables, which would be a fascinating new representation of these familiar objects. It may also be interesting to explore whether these variables shed light on the structure of cuts in parametric space \cite{Britto:2023rig}  or IBPs for Feynman integrals in parametric form \cite{Artico:2023jrc}. The set of subtracted integrals here canonically associate different divergences with subtraction terms, and could perhaps be related to a convenient choice of basis.
\paragraph*{\bf Acknowledgments } 
We thank Nima Arkani-Hamed and Sebastian Mizera for many valuable discussions and collaboration on this subject. We also would like to thank Erik Panzer and Giulio Salvatori for valuable discussions on the topic of subtraction schemes. We also thank Francis Brown, Michi Borinsky, Hofie Hannesdottir, Song He,  and Sebastian Mizera for comments on the draft.

\nocite{*}
\bibliographystyle{JHEP}
\bibliography{references}

\providecommand{\href}[2]{#2}\begingroup\raggedright\begin{thebibliography}{10}

\bibitem{nilsson2010mellin}
L.~Nilsson and M.~Passare, \emph{Mellin transforms of multivariate rational
  functions},  \href{https://arxiv.org/abs/1010.5060}{{\ttfamily 1010.5060}}.

\bibitem{Brown:2019wna}
F.~Brown and C.~Dupont, \emph{{Single-valued integration and superstring
  amplitudes in genus zero}},
  \href{http://dx.doi.org/10.1007/s00220-021-03969-4}{\emph{Commun. Math.
  Phys.} {\bfseries 382} (2021) 815--874},
  [\href{https://arxiv.org/abs/1910.01107}{{\ttfamily 1910.01107}}].

\bibitem{Brown:2009qja}
F.~C.~S. Brown, \emph{{Multiple zeta values and periods of moduli spaces M 0 ,n
  ( R )}}, {\emph{Annales Sci. Ecole Norm. Sup.} {\bfseries 42} (2009) 371},
  [\href{https://arxiv.org/abs/math/0606419}{{\ttfamily math/0606419}}].

\bibitem{Brown:2015fyf}
F.~Brown, \emph{{Feynman amplitudes, coaction principle, and cosmic Galois
  group}}, \href{http://dx.doi.org/10.4310/CNTP.2017.v11.n3.a1}{\emph{Commun.
  Num. Theor. Phys.} {\bfseries 11} (2017) 453--556},
  [\href{https://arxiv.org/abs/1512.06409}{{\ttfamily 1512.06409}}].

\bibitem{Arkani-Hamed:2022cqe}
N.~Arkani-Hamed, A.~Hillman and S.~Mizera, \emph{{Feynman polytopes and the
  tropical geometry of UV and IR divergences}},
  \href{http://dx.doi.org/10.1103/PhysRevD.105.125013}{\emph{Phys. Rev. D}
  {\bfseries 105} (2022) 125013},
  [\href{https://arxiv.org/abs/2202.12296}{{\ttfamily 2202.12296}}].

\bibitem{Schultka:2018nrs}
K.~Schultka, \emph{{Toric geometry and regularization of Feynman integrals}},
  \href{https://arxiv.org/abs/1806.01086}{{\ttfamily 1806.01086}}.

\bibitem{Koba:1969kh}
Z.~Koba and H.~B. Nielsen, \emph{{Manifestly crossing invariant parametrization
  of n meson amplitude}},
  \href{http://dx.doi.org/10.1016/0550-3213(69)90071-6}{\emph{Nucl. Phys. B}
  {\bfseries 12} (1969) 517--536}.

\bibitem{Arkani-Hamed:2019mrd}
N.~Arkani-Hamed, S.~He and T.~Lam, \emph{{Stringy canonical forms}},
  \href{http://dx.doi.org/10.1007/JHEP02(2021)069}{\emph{JHEP} {\bfseries 02}
  (2021) 069}, [\href{https://arxiv.org/abs/1912.08707}{{\ttfamily
  1912.08707}}].

\bibitem{Arkani-Hamed:2019plo}
N.~Arkani-Hamed, S.~He, T.~Lam and H.~Thomas, \emph{{Binary Geometries,
  Generalized Particles and Strings, and Cluster Algebras}},
  \href{https://arxiv.org/abs/1912.11764}{{\ttfamily 1912.11764}}.

\bibitem{He:2020onr}
S.~He, Z.~Li, P.~Raman and C.~Zhang, \emph{{Stringy canonical forms and binary
  geometries from associahedra, cyclohedra and generalized permutohedra}},
  \href{http://dx.doi.org/10.1007/JHEP10(2020)054}{\emph{JHEP} {\bfseries 10}
  (2020) 054}, [\href{https://arxiv.org/abs/2005.07395}{{\ttfamily
  2005.07395}}].

\bibitem{Chetyrkin:1980pr}
K.~G. Chetyrkin, A.~L. Kataev and F.~V. Tkachov, \emph{{New Approach to
  Evaluation of Multiloop Feynman Integrals: The Gegenbauer Polynomial x Space
  Technique}},
  \href{http://dx.doi.org/10.1016/0550-3213(80)90289-8}{\emph{Nucl. Phys. B}
  {\bfseries 174} (1980) 345--377}.

\bibitem{AHHM}
N.~Arkani-Hamed, A.~Hillman and S.~Mizera, \emph{in preparation}, .

\bibitem{Britto:2023rig}
R.~Britto, \emph{{Generalized Cuts of Feynman Integrals in Parameter Space}},
  \href{http://dx.doi.org/10.1103/PhysRevLett.131.091601}{\emph{Phys. Rev.
  Lett.} {\bfseries 131} (2023) 091601},
  [\href{https://arxiv.org/abs/2305.15369}{{\ttfamily 2305.15369}}].

\bibitem{Artico:2023jrc}
D.~Artico and L.~Magnea, \emph{{Integration-by-parts identities and
  differential equations for parametrised Feynman integrals}},
  \href{https://arxiv.org/abs/2310.03939}{{\ttfamily 2310.03939}}.

\bibitem{Borinsky:2023jdv}
M.~Borinsky, H.~J. Munch and F.~Tellander, \emph{{Tropical Feynman integration
  in the Minkowski regime}},
  \href{http://dx.doi.org/10.1016/j.cpc.2023.108874}{\emph{Comput. Phys.
  Commun.} {\bfseries 292} (2023) 108874},
  [\href{https://arxiv.org/abs/2302.08955}{{\ttfamily 2302.08955}}].

\bibitem{Borinsky:2020rqs}
M.~Borinsky, \emph{{Tropical Monte Carlo quadrature for Feynman integrals}},
  \href{https://arxiv.org/abs/2008.12310}{{\ttfamily 2008.12310}}.

\bibitem{Kompaniets:2016hct}
M.~Kompaniets and E.~Panzer, \emph{{Renormalization group functions of $\phi^4$
  theory in the MS-scheme to six loops}},
  \href{http://dx.doi.org/10.22323/1.260.0038}{\emph{PoS} {\bfseries LL2016}
  (2016) 038}, [\href{https://arxiv.org/abs/1606.09210}{{\ttfamily
  1606.09210}}].

\bibitem{Kompaniets:2017yct}
M.~V. Kompaniets and E.~Panzer, \emph{{Minimally subtracted six loop
  renormalization of $O(n)$-symmetric $\phi^4$ theory and critical exponents}},
  \href{http://dx.doi.org/10.1103/PhysRevD.96.036016}{\emph{Phys. Rev. D}
  {\bfseries 96} (2017) 036016},
  [\href{https://arxiv.org/abs/1705.06483}{{\ttfamily 1705.06483}}].

\bibitem{Bergere1974}
M.~C. Berg{\`e}re and J.~B. Zuber, \emph{{Renormalization of Feynman amplitudes
  and parametric integral representation}},
  \href{http://dx.doi.org/10.1007/BF01646611}{\emph{Comm. Math. Phys.}
  {\bfseries 35} (Jun, 1974) 113--140}.

\bibitem{Panzer:2019yxl}
E.~Panzer, \emph{{Hepp's bound for Feynman graphs and matroids}},
  \href{https://arxiv.org/abs/1908.09820}{{\ttfamily 1908.09820}}.

\bibitem{Binoth:2000ps}
T.~Binoth and G.~Heinrich, \emph{{An automatized algorithm to compute infrared
  divergent multiloop integrals}},
  \href{http://dx.doi.org/10.1016/S0550-3213(00)00429-6}{\emph{Nucl. Phys. B}
  {\bfseries 585} (2000) 741--759},
  [\href{https://arxiv.org/abs/hep-ph/0004013}{{\ttfamily hep-ph/0004013}}].

\bibitem{Borowka:2015mxa}
S.~Borowka, G.~Heinrich, S.~P. Jones, M.~Kerner, J.~Schlenk and T.~Zirke,
  \emph{{SecDec-3.0: numerical evaluation of multi-scale integrals beyond one
  loop}}, \href{http://dx.doi.org/10.1016/j.cpc.2015.05.022}{\emph{Comput.
  Phys. Commun.} {\bfseries 196} (2015) 470--491},
  [\href{https://arxiv.org/abs/1502.06595}{{\ttfamily 1502.06595}}].

\bibitem{Nilsson2013}
L.~Nilsson and M.~Passare, \emph{Mellin transforms of multivariate rational
  functions}, \href{http://dx.doi.org/10.1007/s12220-011-9235-7}{\emph{Journal
  of Geometric Analysis} {\bfseries 23} (Jan, 2013) 24--46},
  [\href{https://arxiv.org/abs/1010.5060}{{\ttfamily 1010.5060}}].

\bibitem{YelleshpurSrikant:2019khx}
A.~Yelleshpur~Srikant, \emph{{Spherical Contours, IR Divergences and the
  geometry of Feynman parameter integrands at one loop}},
  \href{http://dx.doi.org/10.1007/JHEP07(2020)236}{\emph{JHEP} {\bfseries 07}
  (2020) 236}, [\href{https://arxiv.org/abs/1907.05429}{{\ttfamily
  1907.05429}}].

\bibitem{speer1969generalized}
E.~Speer, \emph{Generalized Feynman Amplitudes}.
\newblock Annals of Mathematics Studies. Princeton University Press, 1969.

\bibitem{Itzykson:1980rh}
C.~Itzykson and J.~Zuber, \emph{{Quantum Field Theory}}.
\newblock International Series In Pure and Applied Physics. McGraw-Hill, New
  York, 1980.

\bibitem{Becher:2009qa}
T.~Becher and M.~Neubert, \emph{{On the Structure of Infrared Singularities of
  Gauge-Theory Amplitudes}},
  \href{http://dx.doi.org/10.1088/1126-6708/2009/06/081}{\emph{JHEP} {\bfseries
  06} (2009) 081}, [\href{https://arxiv.org/abs/0903.1126}{{\ttfamily
  0903.1126}}].

\bibitem{Collins:1989gx}
J.~C. Collins, D.~E. Soper and G.~F. Sterman, \emph{{Factorization of Hard
  Processes in QCD}},
  \href{http://dx.doi.org/10.1142/9789814503266_0001}{\emph{Adv. Ser. Direct.
  High Energy Phys.} {\bfseries 5} (1989) 1--91},
  [\href{https://arxiv.org/abs/hep-ph/0409313}{{\ttfamily hep-ph/0409313}}].

\bibitem{Becher:2014oda}
T.~Becher, A.~Broggio and A.~Ferroglia, \emph{{Introduction to Soft-Collinear
  Effective Theory}}, vol.~896.
\newblock Springer, 2015,
  \href{http://dx.doi.org/10.1007/978-3-319-14848-9}{10.1007/978-3-319-14848-9}.

\bibitem{brown2011multiple}
F.~Brown, \emph{{Multiple zeta values and periods: from moduli spaces to
  Feynman integrals}}, {\emph{Contemp. Math} (2011) 27--52}.

\bibitem{BrownIHES}
F.~Brown, ``{Motivic periods and the cosmic Galois group (IHES, May 2015)}.''

\bibitem{PhysRev.118.838}
S.~Weinberg, \emph{{High-Energy Behavior in Quantum Field Theory}},
  \href{http://dx.doi.org/10.1103/PhysRev.118.838}{\emph{Phys. Rev.} {\bfseries
  118} (May, 1960) 838--849}.

\bibitem{Hepp1966}
K.~Hepp, \emph{{Proof of the Bogoliubov-Parasiuk theorem on renormalization}},
  \href{http://dx.doi.org/10.1007/BF01773358}{\emph{Comm. Math. Phys.}
  {\bfseries 2} (Dec, 1966) 301--326}.

\bibitem{Zimmermann1969}
W.~Zimmermann, \emph{{Convergence of Bogoliubov's method of renormalization in
  momentum space}}, \href{http://dx.doi.org/10.1007/BF01645676}{\emph{Comm.
  Math. Phys.} {\bfseries 15} (Sep, 1969) 208--234}.

\end{thebibliography}\endgroup

\end{document}